\documentstyle[aps,prl,epsf,floats]{revtex}

\def\ep{\epsilon_{\perp}}
\def\sp{\sigma_{\perp}}
\def\tp{t_{\perp}}
\def\a{\alpha}

\def\o{\omega}
\def\iomn{i\omega_n}
\def\R{\mbox{Re}}
\def\X{\mbox{X}}
\def\K{\mbox{K}}
\def\kro{K_{\rho}}
\def\est{E^*}
\def\cm1{\mbox{cm}^{-1}}
\def\SCl{$(\mbox{TMTSF})_2\mbox{ClO}_4$ }
\def\SP{$(\mbox{TMTSF})_2\mbox{PF}_6$ }

\def\SX{$(\mbox{TMTSF})_2\mbox{X}$ }

\begin{document}
\twocolumn[\hsize\textwidth\columnwidth\hsize\csname@twocolumnfalse%
\endcsname
\title{Interchain conductivity of coupled Luttinger liquids and
organic conductors}
\author{Antoine Georges${}^{1,2}$, Thierry Giamarchi${}^{2,1}$,
Nancy Sandler${}^{3}$}
\address{${}^{1}$CNRS UMR 8549- Laboratoire de Physique Th{\'e}orique, Ecole Normale
Sup{\'e}rieure, 24 Rue Lhomond 75005 Paris FRANCE\\ ${}^{2}$
Laboratoire de Physique des Solides, CNRS-UMR 8502, Batiment 510,
Universit\'e de Paris-Sud, 91405 Orsay, FRANCE \\ ${}^{3}$ Laboratoire de
Physique de la Matiere Condens\'ee, Ecole Normale Sup{\'e}rieure, 24
Rue Lhomond 75005 Paris FRANCE}
\date{\today}
\maketitle

\begin{abstract}
We reconsider the theory of dc and ac interchain conductivity
in quasi-one dimensional systems. Our results are in good agreement
with the measured c-axis optical
conductivity of \SCl and suggest that the c-axis
dc-conductivity of \SP in the $150K < T < 300K$ range is dominated
by precursor effects of Mott localization.
The crossover from a Luttinger liquid at high energy to a Fermi liquid
at low energy is also addressed, within a dynamical mean-field theory.
Implications for the inter-chain resistivity and Drude weight in
the Fermi liquid regime are discussed.
\end{abstract} \vspace{.2cm}
]

Inter-chain (or inter-plane) coherence is a key issue for
the physics of anisotropic strongly correlated systems. It is of
special importance for the quasi-one dimensional organic conductors
TMTSF$_{2}$X and TMTTF$_{2}$X which are systems of weakly coupled
chains \cite{jerome_organic_review}. Since the one dimensional (1D)
interacting  electron gas has non Fermi liquid (Luttinger liquid)
properties \cite{schulz_houches_revue}, one expects in these compounds
a crossover from Luttinger liquid behavior to a coherent behavior as
the temperature or frequency is lowered. The latter can be either a
metallic three-dimensional conductor (presumably Fermi-liquid like), or
a phase with long-range order.

Although there is general agreement
on this qualitative scenario, there is considerable debate
on how precisely this crossover takes place
\cite{bourbonnais_couplage,wen_coupled_chains,yakovenko_manychains,%
clarke_coherence_coupled,schulz_moriond}. As was pointed out early on
\cite{bourbonnais_rmn}, the interactions renormalize
the effective interchain hopping of weakly-coupled Luttinger
chains. This leads to a crossover scale $\est \propto t(\tp/t)^{1/(1-\a)}
= \tp(\tp/t)^{\a/(1-\a)}$
for the onset of coherence along one of the transverse directions with
interchain hopping $\tp$. In this expression, $t$ is the intra-chain
hopping and $\a\equiv {1\over 4}(\kro+1/\kro)-{1\over 2}$ is the Fermi
surface exponent. The scale $\est$ can be much smaller than the bare
interchain hopping $\tp$. With this idea in mind, analysis of NMR data
\cite{bourbonnais_rmn_2,wzietek_nmr} led to believe
that the 1D Luttinger liquid (LL) to Fermi liquid (FL) crossover
took place in the \SX family at a low temperature, of the
order of $10-15\K$. This scale seemed consistent with the above
estimate of $\est$ since the interchain hopping along the $b$-axis
is $t_b\sim 300K$, an order of magnitude smaller than the hopping $t_a$ along
the chains. Recent analysis of the transport and optical properties
\cite{giamarchi_mott_shortrev} have led to a reexamination of this
commonly accepted view. Measurement of the optical conductivity along
the chains agrees with a
LL picture at high enough frequency and provides a
determination of the LL exponent $\kro$
\cite{dressel_optical_tmtsf,schwartz_electrodynamics} (of order
$\kro\simeq 0.23$). This value is consistent with the temperature
dependence of $\rho_a(T)$ in the $100\K-300\K$ range (suitably corrected for thermal expansion
\cite{jerome_organic_review}). However, these studies
and most notably the recent measurements of the interchain dc conductivity
\cite{moser_conductivite_1d} revealed that
the 1D regime only extends down to $\sim 100\K$, a much
higher temperature scale than the early estimates based on the NMR.
Furthermore, it
is clear from the optical data that another important energy scale is
present in these compounds, associated with a Mott gap signaled by the
onset of optical absorption at $\sim 100\cm1\simeq 150\K$.
Transverse hopping is actually responsible for the metallic character of
the \SX compounds, which otherwise would be Mott insulators
\cite{giamarchi_mott_shortrev,vescoli_confinement_science}.
These findings raise a number of important and largely
unanswered questions such as the precise nature of the dimensional
crossover, the origin of the anomalies observed for $10K < T <100K$
(calling in particular for a reinterpretation of the NMR in this
range), and the actual range of the LL
behavior.

In this letter, we focus on
{\it inter-chain} transport and optical response, which are ideal
probes of these issues.
We first reconsider the theoretical analysis of the
$\omega$ and $T$- dependence of the inter-chain
conductivity for the simplest case of weakly coupled metallic Luttinger
liquids. We find a power law of the temperature or the frequency
$\sigma_\perp(T,\omega)\propto (T,\omega)^{2\alpha-1}$. The exponent is
at variance with previous theoretical predictions
\cite{clarke_coherence_coupled,anderson_quasi1d_coherence,%
moser_conductivite_1d}, but is in reasonable agreement with recent
measurements of c-axis optical conductivity of the \SCl compound
\cite{henderson_transverse_optics_organics}.
We then show that understanding the observed T-dependence of c-axis
transport in \SP requires to take into
account the physics associated with the Mott gap
apparent in optical measurements. Finally, we consider a dynamical
mean-field theory (DMFT) for the description of the dimensional crossover,
and draw some consequences for the inter-chain conductivity in the
low-temperature FL regime.

We consider the hamiltonian of coupled chains:
\begin{equation}
\label{ham}
H = \sum_{m}\,H_{1D}^{(m)} - \sum_{\langle m,m'\rangle}\tp \,
\sum_{i\sigma} (c^+_{im\sigma}c_{im'\sigma} + \mbox{h.c})
\end{equation}
in which $H_{1D}^{(m)}$ stands for each individual chain
$m$, and the inter-chain hopping $\tp$
will be first considered, for simplicity, to be identical for all
transverse directions. Using the DMFT approach detailed below, a
general formula (Eq.~\ref{sigmatrue}) for the transverse conductivity
can be established. When the inter-chain
hopping $t_\perp$ can be treated in perturbation theory to lowest
order, this formula reduces to that
established in Ref.\cite{moser_conductivite_1d}
within the tunneling approximation
\cite{cond-units}:
\begin{eqnarray}
\nonumber
\R\,\sp (\o,T)^{pert}\propto
&&\tp^2\int dx\int d\omega'
A_{1D}(x,\o+\o')\times\\
&&\times A_{1D}(x,\o'){{f(\o')-f(\o'+\o)}\over{\o}}
\label{sigmapert}
\end{eqnarray}
where $A_{1D}(x,\o)$ is the one-electron spectral function of a single
chain, and $f$ the Fermi function. Let us consider first the simplest case in
which each decoupled chain is a Luttinger liquid characterized by a
given value of $\kro$, and all intra-chain umklapp processes can be
neglected. For the values of $\kro$ of interest here, the transverse hopping is
then a relevant perturbation in the renormalization group sense,
so that it is legitimate to treat it
perturbatively only at sufficiently high energy or temperature.
Eq.~(\ref{sigmapert}) thus applies only when
$kT\gg\est$ or $\hbar\o \gg \est$, where $\est = t (\tp/t)^{1/(1-\a)}$
is the single-particle crossover scale mentioned above
\cite{bourbonnais_rmn,bourbonnais_couplage,boies_couplage}.
In this high-temperature/high-energy regime, the system can be considered to be
quasi one-dimensional. Using known properties of the spectral functions in a
Luttinger liquid, we then obtain from (\ref{sigmapert}):
\begin{equation} \label{eq:simple}
\label{sigma_1D} \R\,\sp (\o,T)^{pert}\propto
\left(\frac{\tp}{t}\right)^2 \left(\frac{T}{t}\right)^{2\a-1}
F_\alpha\left({{\o}\over{T}}\right)
\end{equation}
In this expression, $F_\alpha$ is a scaling function such that
$F_\alpha(0)=\mbox{const}$ and $F_\alpha(x\gg 1)\sim x^{2\a-1}$. Thus, we find
$\R \sp (T\gg\o) \propto T^{2\a-1}$ for $kT\gg\est$, while
at high-frequency ($\hbar\o\gg\est$), we obtain
$\sp(\o\gg T)\propto \o^{2\a-1}$.

This behavior of $\sp(T)$ differs from the conclusions of previous
authors. In \cite{clarke_coherence_coupled}, the authors used a
somewhat different approach than the direct Kubo formula and concluded
that $\sp(T)\propto T^{2\a}$. In \cite{moser_conductivite_1d}, it was
noted that this temperature dependence also results from
(\ref{sigmapert}) {\it provided} that the spatial integral is cutoff at
a scale of the order of the lattice spacing. According to these
authors, this cutoff is justified by the incoherent nature of the
propagation of a physical electron within each chain (in which the true
quasiparticles are spinons and holons). In our opinion however, this
information is already encoded in the spectral function of the physical
electron $A_{1D}(x,\o)$ and no extra cutoff has to be introduced
in (\ref{sigmapert}). This conclusion would of course be changed in the
presence of specific forms of intra-chain disorder such as a random
forward scattering, which would spoil momentum conservation and bring
in a natural cutoff of the order of the mean-free path. Note however
that more realistic forms of disorder would be signaled by
localisation effects.

We now turn to a comparison with experiments on $(\mbox{TMTSF})_2\X$.
It is increasingly recognized
that the properties of these compounds result from the commensurate
{\it quarter-filling} of the
band \cite{giamarchi_mott_shortrev,schwartz_electrodynamics,moser_conductivite_1d}.
Transport and optical measurements lead to consistent values of the LL parameter
$\kro \sim 0.23$. Note that $\kro < 0.25$ is the condition upon which the
quarter-filled umklapp process becomes relevant. This suggests
that if the transverse hopping was entirely suppressed in these compounds, they
would actually be {\it one-dimensional (quarter filled) Mott insulators}.
Fortunately, optical experiments
extend to high frequencies compared to the Mott gap. Thus
the complications associated with the proximity to the Mott insulating
state can be treated perturbatively, making the
comparison to Eq.(\ref{eq:simple}) relevant. Recently, optical
studies of \SCl have been performed
\cite{henderson_transverse_optics_organics}, and the high-frequency
optical conductivity $\sigma_c(\o)$ along the {\it least conducting}
c-axis was found to be a broad, very slowly increasing function of
frequency in the range $50\cm1 <\o < 10^5\cm1$. This slowly
increasing continuum is essentially independent of temperature
for $10\K <T< 300 \K$,
and is in good agreement with our result $\sigma_c(\o) \propto \o^{2\a-1}$.
The exponent $2\a-1=(\kro+1/\kro)/2-2$ depends quite
sensitively on the value of $\kro$ ($2\a-1 = 0.12$ for $\kro=0.25$
and $2\a-1=0.6$ for $\kro=0.2$). The latter value $2\a-1\simeq 0.6$
provides a reasonable fit to the experimental
data of \cite{henderson_transverse_optics_organics}.
Note that values of $\kro$ in the $0.2$-$0.25$ range are quite consistent with the other
experimental findings discussed above. In contrast, the $\omega^{2\a}$
law advocated in \cite{clarke_coherence_coupled,anderson_quasi1d_coherence,%
moser_conductivite_1d} would require
$\kro\simeq 0.35$ to explain these data, a value quite inconsistent with
that obtained from the longitudinal optical conductivity.
Finally, an important issue in view of the success of this
simple theory is whether it can also account for the behavior of the
optical conductivity along the b-axis. Since $t_b/t_a\simeq 0.1$, an
incoherent regime should also apply there at those high frequencies.
Preliminary analysis \cite{degiorgi_private_sigmab} of the
$\sigma_b$ data of \cite{henderson_transverse_optics_organics} seems
in agreement with such a behavior, but more detailed
comparison with the data is needed.

We now discuss the measurements of the c-axis transport in \SP
\cite{moser_conductivite_1d,moser_thesis}. Because of the large thermal expansion,
these were performed at several pressures
so that a volume correction could be made. This yields data
for the c-axis resistivity $\rho_c(T)$ which can be interpreted as
corresponding to a compound with constant lattice parameters.
This volume-corrected $\rho_c(T)$ displays a marked
increase by more than a factor of $3$ upon cooling from
$T=350 \K$ down to $T= 100\K$ (followed by a rapid downturn at
lower temperature which presumably marks the onset of inter-chain
coherence). Such a large variation of $\rho_c(T)$
{\it cannot be explained} by the dependence $\rho_c\propto 1/T^{2\a-1}$
found above with acceptable values of $\kro$ (which correspond to rather
small values of $2\a-1$).
In \cite{moser_conductivite_1d}, it was fitted to a $1/T^{1.4}$ power law,
which happens to be in reasonable agreement with the $1/T^{2\a}$ dependence
advocated by these authors. However, the
temperature range in which these data are obtained is such that
$kT$ is never significantly larger than the ``would-be'' 1D Mott gap
which can be extracted from the optical measurements
($100-200\cm1$) leading to transport gaps (half the optical gap)
in the range $\Delta \simeq 75-150\K$.
As a result, the proximity to the incipient Mott insulator cannot be neglected,
and we propose that this effect is the one responsible
for the increase of $\rho_c(T)$. Indeed, we have attempted a
phenomenological fit by an
{\it activated behavior} $\rho_c(T)\propto 1/T^{2\a-1}\exp{\Delta/T}$.
and found that it provides a fit of the data of
Ref.\cite{moser_conductivite_1d,moser_thesis}
which is as satisfactory as the power-law fit used
in \cite{moser_conductivite_1d}.
The value of $\Delta$ used in this fit is of the same order of magnitude
than the Mott gap seen in optical measurements. The discrepancy between the
theoretical result, Eq.(\ref{sigma_1D}) and the experimentally observed
behavior of $\rho_c(T)$ is, in our opinion, a strong indication that an explanation
based on incoherent tunneling between LL chains
in a {\it purely metallic} regime is insufficient and that the physics
of Mott localisation plays an important role in the temperature range
$150\K <T <300 \K$.

A more refined theory of the interplay between interchain hopping  and
Mott localisation is clearly needed, including both the transverse hopping
and the quarter- filling umklapp scattering \cite{florens_transverse_long}.
Due to the umklapp, the effective $\kro$ will be renormalized towards smaller
values as temperature is decreased, hence leading to larger values of
$\rho_c$, as expected. The main issue is whether the
{\it intra-chain} resistivity is also strongly affected by these
renormalizations. Indeed, no sign
of an upturn in $\rho_a(T)$ is observed experimentally for the
$\mbox{TMTSF}_2X$ compounds. Since
$\rho_a(T) \propto g_{1/4}^2\,T^{16\kro-3}$ \cite{giamarchi_mott_shortrev},
it is conceivable that the decrease in $\kro$ is almost compensated by an associated
decrease of the effective $g_{1/4}$ due to $\tp$.
This however, remains to be verified and is probably the most
challenging issue for the picture that we propose.
Transport in the $\mbox{TMTTF}_2X$ family is interesting in
this respect. These compounds are insulators with a rather
large Mott gap, and indeed both
$\rho_a(T)$ and $\rho_c(T)$ display an activated behavior at low temperature.
Above some temperature scale however, $\rho_a(T)$ becomes metallic, while
$\rho_c(T)$ still displays insulating behaviour (albeit slower than activated).
Studies of the transport properties of these compounds over a more extended
range of pressure and temperature (and in particular a detailed comparison of
the energy scales appearing in $\sigma_a(\omega)$, $\rho_a(T)$ and $\rho_c(T)$)
would certainly prove very interesting.

Let us finally address the issue of the crossover between a
Mott-Luttinger liquid at high energy/temperature and a coherent
(presumably Fermi-liquid) state at low energy. The main
theoretical difficulty is that this crossover is associated with
the breakdown of perturbative expansions in $\tp$.
On the other hand, perturbative approaches in the
interaction strength could be used in the FL regime, but fail in
the LL regime. The necessity of treating $\tp$ in a
non-perturbative manner was recently noted by Arrigoni
\cite{arrigoni_tperp_resummation}, who used a limit of infinite
transverse dimensionality as a guidance for resumming the
perturbation series in $\tp$. Here, we show that this limit
provides a generalized dynamical mean-field theory (DMFT)
framework for the description of the crossover, and we draw some
consequences for inter-chain conductivity. The limit of infinite
dimensions has been used extensively in recent years to formulate
a controlled DMFT of strongly-correlated systems \cite{georges_d=infini,additional_large_d}.
We consider each one-dimensional chain to be
coupled to $z_{\perp}\rightarrow\infty$ nearest neighbor ones,
with the transverse hopping scaled as
$\tp=\widetilde{t}_\perp/\sqrt{z_{\perp}}$, so that the
non-interacting density of states in the transverse direction
$D(\epsilon_\perp)\equiv\sum_{k_\perp}\delta[\epsilon_\perp-
\epsilon(k_\perp)]$ remains well-defined. It is easily shown that in this limit the
self-energy becomes {\it independent of transverse momentum
$k_\perp$}: $\Sigma=\Sigma(k,\iomn)$ ($k$ is the momentum along
the chains). Furthermore, the calculation of $\Sigma$ reduces to
the solution of an effective one-dimensional problem specified by
the effective action:
\begin{eqnarray}
\nonumber
S_{\text{eff}}=&&
- \int\int^{\beta}_{0}d\tau\,d\tau' \sum_{ij,\sigma}
c^{+}_{i\sigma}(\tau) {\cal G}_0^{-1}(i-j,\tau-\tau')
c_{j\sigma}(\tau')\\
&&+\int_0^{\beta} d\tau H^{int}_{1D}[\{c_{i\sigma},c^+_{i\sigma}\}]
\label{Seff}
\end{eqnarray}
where $H^{int}_{1D}$ is the interacting part of the
on-chain hamiltonian in (\ref{ham}), while ${\cal G}_0$ is an
{\it effective bare propagator} which is determined from a
self-consistency condition. The latter imposes that the
Green's function
$G(i-j,\tau-\tau')\equiv -\langle c(i,\tau)c^+(j,\tau')
\rangle_{\text{eff}}$
calculated from $S_{\text{eff}}$
coincides with the on-chain Green's function of the original
problem, with a  self-energy $\Sigma={\cal G}_0^{-1}-G^{-1}$.
This condition reads:
\begin{equation}\label{sc_cond}
G(k,\iomn) = \int d\ep
{{D(\ep)}\over{\iomn+\mu-\epsilon_k-\Sigma(\iomn,k)-\ep}}
\end{equation}
The DMFT equations (\ref{Seff},\ref{sc_cond})
fully determine the self-energy and Green's function of the
coupled chains. Furthermore, inter-chain transport properties
simplify in this approach. Because the current vertex is
an {\it odd} function of the transverse momentum,
vertex corrections drop out of the inter-chain conductivity, which reads:
\begin{eqnarray}
\nonumber
\R\,\sp (\o,T) \propto
&&\tp^2\int d\ep D(\ep)
\int {{dk}\over{2\pi}}\int d\omega'A(\ep,k,\o')\\
&&\times A(\ep,k,\o+\o'){{f(\o')-f(\o'+\o)}\over{\o}}
\label{sigmatrue}
\end{eqnarray}
Here, $A(\ep,k,\o)=-{1\over\pi}\mbox{Im}G(\ep,k,\o)$ is the
single-particle spectral function of the coupled chains system.
Eq.(\ref{sigmatrue}) has a much wider range of applicability than its
small-$\tp$ limit, Eq.(\ref{sigmapert}), since it holds all the way from
the LL to the FL regime.

Solving quantitatively these DMFT equations is still a challenging
problem however. In the single-site DMFT (corresponding to the usual
$d=\infty$ limit \cite{georges_d=infini}) the mapping is onto a {\it single-impurity}
Anderson model with a self-consistent bath, and several techniques have
been developed to handle this problem. Here however, the
mapping is onto a self-consistent {\it 1D-chain}, and the problem is
an order of magnitude more difficult.
Nevertheless, several conclusions can be drawn from the above equations
even in the absence of a full solution. We shall consider again the simplest
case of metallic 1D LL chains, neglecting umklapp
scattering and Mott physics. In the 1D regime, the self-energy
behaves as $\Sigma\propto t ((\omega,k)/t)^{1/(1-\a)}$. From
(\ref{sc_cond}), it is clear that the inter-chain hopping becomes relevant
when $\tp > \Sigma$, and we recover the crossover
scale $\est\propto t (\tp/t)^{1/(1-\a)}$.
At energies smaller than $\est$, the DMFT approach leads to
Fermi-liquid behavior. This is clear from viewing the effective 1D
model as a set of Anderson impurity models coupled together by the non-local
part of the ``bare'' propagator ${\cal G}_0$. Each of the decoupled
Anderson model is a Fermi-liquid at low-energy and the non-local
hybridization is a smooth perturbation on the FL ground state. In
other words, the physics below $\est$ is smoothly
connected from small $\tp/t$ to large $\tp/t$.
When $\tp\ll t$ and for energy
scales much smaller than the bandwidth (of order $t$),
all single-particle quantities should obey a scaling behavior in
the variables $\tilde{\omega}=\omega/\est$, $\tilde{x}=x\est/t$, $\tilde{T}=T/\est$.
In particular
$t\Sigma(x,\o,T) = \est\tp\tilde{\Sigma}(\tilde{x},\tilde{\omega},\tilde{T})$,
$t G(x,\o,T) = (\est/\tp)\tilde{G}(\tilde{x},\tilde{\omega},\tilde{T})$
with $\tilde{\Sigma}$ and $\tilde{G}$ universal scaling functions
associated with the crossover.
A low-frequency expansion of the scaling form of $\Sigma$ in the FL regime
yields a finite quasiparticule residue $Z\propto (\tp/t)^{{{\a}\over{1-\a}}}
\propto \est/\tp$ (shown in \cite{arrigoni_tperp_resummation} to have only
weak k-dependence along the Fermi surface).
$Z$ can be much smaller than unity, while the effective mass (specific heat)
ratio $m^*/m$ does not become large because $\partial\Sigma/\partial\omega$
and $\partial\Sigma/\partial k$ scale in the same manner.
Using the FL form of $\Sigma$ in
(\ref{sigmatrue}), we obtain the $T=0$ inter-chain conductivity as:
$\sigma_\perp(\omega,T=0)={{\omega^2_{D\perp}}\over{8}}\delta(\o)
+ \sigma_\perp^{reg}(\o)$ with
${\omega_{D\perp}}^2/8 \propto Z^2\tp^2/t = t (\tp/t)^{2/(1-\a)}$.
The total spectral weight
$\int d\o \sigma_\perp(\o)= \omega^2_{P\perp}/8$ is,
from the f-sum rule,
proportional to the inter-chain kinetic energy, and hence to $\tp^2/t$, so that
the relative weight carried by the Drude peak is small, of order
$\omega^2_{D\perp}/\omega^2_{P\perp}\propto (\tp/t)^{2\alpha/(1-\a)} = Z^2$
This is in reasonable agreement with the optical data on the b-axis
reflectivity of \SP \cite{vescoli-thesis}, for which a Drude peak is seen
at low temperature,
with $\omega^2_{D\perp}/\omega^2_{P\perp}\simeq 0.03$
(while $t_b/t_a\simeq 0.1$). Scaling considerations also lead to
an inter-chain resistivity
$\rho_\perp(T)/\rho_0 = (t/\est) R(T/\est)$, with
$R(x\gg 1)\sim x^{1-2\a}$ , $R(x\ll 1)\sim x^2$ and
$\rho_0 = hV_m/(e^2a^2_\perp)$ \cite{cond-units}. This implies that
$\rho_\perp/\rho_0 = A (T/t)^2$ in the FL regime, with an enhanced value of
the coefficient $A\propto (t/\tp)^{{3}\over{1-\a}}$. Note that $\rho_\perp$
is typically much bigger than the Mott limit $\rho_0$, as observed experimentally.

In conclusion, we have reconsidered in this paper the theory of
interchain conductivity in quasi-1D systems. When applied to organic
compounds, good agreement is found with the frequency dependence of
$\sigma_c(\o)$. while c-axis dc transport appears to be dominated by
the proximity of the Mott gap $\Delta$. This calls for additional
theoretical work in the regime where $\est\propto t (\tp/t)^{1/(1-\a)}
< \Delta <\tp$. It would also be valuable to see whether the regime
$\Delta < \est$, which is simpler to analyze theoretically, can be
realized experimentally in some family of compounds (e.g by applying
uniaxial stress to TMTTF$_2$X).

We thank D. Jerome, J. Moser, L. Degiorgi, G.
Gr\"{u}ner, and V. Vescoli for sharing and discussing their
data with us. Discussions with M. Gabay and A. M Tremblay are also
acknowledged. A.G. and T.G. would like to thank the ITP, Santa Barbara
for support (under NSF grant No. PHY94-07194) and hospitality during
part of this work.


\begin{thebibliography}{10}
\bibitem{cond-units}
Transverse conductivities are given in the units of $e^2a^2_\perp/(hV_m)$, with
  $a_\perp$ the inter-chain spacing and $V_m$ the volume of the unit cell, and
  the on-chain lattice spacing is normalized to unity.
\end{thebibliography}

\end{document}